\documentclass[10pt, conference, compsocconf]{IEEEtran}

\usepackage{graphicx}
\usepackage{subfig}
\usepackage{tikz}
\usetikzlibrary{shapes.misc, positioning,calc}
\usetikzlibrary{fit}
\usepackage{amsmath}
\usepackage{multirow}
\usepackage{booktabs}
\usepackage{ifthen}
\usepackage{url}
\usepackage{ifthen,color}
\usepackage{moresize}
\usepackage{algorithmic}
\usepackage{color}
\definecolor{darkgreen}{rgb} {0.0,0.5,0.0}
\definecolor{darkblue}{rgb} {0.0,0.0,0.5}
\definecolor{bluegreen}{rgb}{0.0,0.5,0.5}

\definecolor{dr} {rgb} {0.7,0.0,0.0}
\definecolor{lg} {rgb} {0.7,0.3,0.9}
\newcommand{\revision} [1] {#1}
\newcommand{\remove} [1] {#1}
\newcommand{\replace}[2]{#2}

\newcommand{\code}[1]{\textsf{#1}}

\newcommand{\defn}[1]{\ifmmode\text{\emph{\textbf{#1}}}\else\emph{\textbf{#1}}\fi}
\newcommand{\pairs}[2]{\langle #1, #2 \rangle}

\ifdefined\textln\relax\else\newcommand{\textln}[1]{#1}\fi
\ifdefined\shortcite\relax\else\newcommand{\shortcite}[1]{\cite{#1}}\fi
\ifdefined\nd\relax\else\newcommand{\nd}[1]{#1D}\fi

\newlength\myindent
\begin{document}

%
\title{GPU LSM\@: A Dynamic Dictionary Data Structure for the GPU}


\author{%
\IEEEauthorblockN{Saman Ashkiani, Shengren Li}
\IEEEauthorblockA{
University of California, Davis\\
\{sashkiani, shrli\}@ucdavis.edu}
\and
\and
\IEEEauthorblockN{Martin Farach-Colton}
\IEEEauthorblockA{
Rutgers University\\
farach@cs.rutgers.edu}
\and
\IEEEauthorblockN{Nina Amenta, John D. Owens}
\IEEEauthorblockA{
University of California, Davis\\
\{amenta@cs, jowens@ece\}.ucdavis.edu}
}

\maketitle
\begin{abstract}
We develop a \emph{dynamic} dictionary data structure for the 
GPU, supporting fast insertions and deletions, 
based on the Log Structured Merge tree (LSM). 
Our implementation on an NVIDIA K40c GPU has an average update 
(insertion or deletion) rate of 225 M~elements/s, 13.5x faster than merging items into a sorted array. 
The GPU LSM supports the retrieval operations of 
lookup, count, and range query operations with an average rate of 75 M, 32 M and 23 M~queries/s respectively. 
The trade-off for the dynamic updates is that the sorted array is
almost twice as fast on retrievals. 
We believe that our GPU LSM is the first dynamic general-purpose dictionary data structure for the GPU\@.
\end{abstract}

\graphicspath{{}}
\section{Introduction}\label{sec:intro}

GPU programmers generally do not consider applications that require modifications to their data structures.
NVIDIA's OptiX ray tracing engine~\cite{Parker:2010:OAG}, for instance, with very limited exceptions~\cite{Kopta:2012:FEB}, entirely rebuilds its BVH tree  every frame.
This is in stark contrast to the sequential and external memory scenarios,
in which dynamic (mutable) data structures are a cornerstone of programming practice.
Providing concurrency to, and guaranteeing correctness within, a mutable data structure on the massively parallel GPU is a significant challenge.
We believe that dynamic GPU data structures that can be both efficiently built and
updated on the GPU will enlarge the scope of problems that can be addressed, especially as data sizes increase.

In this paper we begin this study by considering a dynamic version of one of the most
basic abstract data types, the dictionary, which supports not only lookups but range and counting queries.
Potential applications that might benefit from efficient dynamic GPU data structures, especially those supporting range queries, include
finding connected components in maximally stable extremal regions (MSER) problem in computer vision; processing dynamic graphs and trees; processing moving objects (e.g., real-time range queries to find $k$ nearest neighbors for all moving objects in a 2D plane~\cite{Lettich:2014:GCR}); processing spatial data (e.g., real-time tweet visualization from a user-defined geographical region~\cite{Mostak:2016:UGA}); and dynamic memory allocators.


\subsection{Batch operations}
Depending on the number of updates required to a dynamic GPU data structure, we might find ourselves in one of three scenarios;
for simplicity, let's just refer to insertion.
If we only need to insert one, or a few, items,
we could fall back on one of many existing serial algorithms.
In this situation, there is not enough work to occupy a modern GPU
In the second scenario,
the number of items to insert is at least on the order of the number of items already in the data structure.
In this case it is likely most efficient to simply rebuild the whole data structure.
The third, and interesting, scenario forms
the very large middle ground between serial insert and complete rebuild,  when there are
more than a few insertions but not on the order of the current size of the data structure.
In this paper we focus on this case in which an
update of the data structure is potentially less expensive than a full rebuild.
%

We define a \emph{batch} operation on the GPU as one where the number
of items to insert or delete is large enough that the operations can profitably be done in parallel, but
not so large that we might as well rebuild the data structure.
We design a data structure to handle
a batch of queries or updates, so that we can exploit concurrency across the batch.
The reason this is a challenge is that by increasing the number of concurrent modifications to a shared data structure, it
becomes harder to cooperate while maintaining correctness.
For example, in a balanced tree data structure,
inserting an item requires rebalancing the tree, and parallel inserts might require rebalancing operations that interfere with each other.
Our challenge is to find or design data structures that support batched operations and to implement efficient algorithms for them.
We will discuss our specific assumptions for the semantics of batch operations
in Section~\ref{subsec:cola_design}.

\subsection{Dictionaries}
\label{sec:dictionaries}
We focus on the dictionary abstract data type, because it is fundamental to many algorithms and implemented in
many interesting ways in the sequential and external-memory contexts.
Formally, a \emph{dictionary}
maintains a set $\mathcal S$ of $\pairs{\defn{key}}{\defn{value}}$ pairs,
where keys are totally ordered, that
supports the following operations:

\begin{itemize}
\item \textsc{insert}$(k,v)$:
$\mathcal S \leftarrow (\mathcal S  -  \{ \pairs{k}{*} \}) \cup \{ \pairs{k}{v}\}$.
If we insert a new item whose key matches an item already
in the set, we replace the old item.

\item \textsc{delete}$(k)$:
  $\mathcal S \leftarrow \mathcal S - \{ \pairs{k}{*}\}$.  Remove all
  key-value pairs with key $k$.

\item \textsc{lookup}$(k)$: return $\pairs{k}{v}\in\mathcal S$, or $\perp$ if no such key exists.

\item \textsc{count}$(k_1,k_2)$: return the number of pairs
  $\langle k,* \rangle$ such that $k_1\leq k \leq k_2$.

\item \textsc{range}$(k_1,k_2)$: returns all pairs
  $\langle k,* \rangle$ such that $k_1\leq k \leq k_2$.

\end{itemize}
Insert and delete are \textit{update} operations that modify the state of the dictionary, while
lookup, count and range queries are \textit{retrieval} operations\footnote{In this work we have only considered \textsc{lookup}, \textsc{count} and \textsc{range} queries. However, it is straightforward to support other order-based queries such as finding a successor or a predecessor of a certain key.}, which are read-only.

Certainly existing GPU data structures, for example arrays,
sorted arrays and hash tables, can implement all of these operations, but not necessarily efficiently.
For example, a sorted array is adequate for all of the retrieval operations.
But updating a sorted array, even by the insertion or deletion of a single element, potentially requires resorting or merging the entire data structure and hence a large number of memory operations.
A hash table~\cite{Alcantara:2009:RPH:nourl} supports even more efficient lookups but because it
is an unordered data structure, count and range queries are impractical, and
existing GPU hash table implementations do not support updates.
A data structure with no support for mutability can (and should!)
outperform a mutable data structure on retrieval operations,
but incremental changes to a mutable data structure can be
considerably cheaper than an entire rebuild of a non-mutable data structure.

The lack of existing mutable data structures on GPUs emphasizes the challenge in identifying and/or designing a data structure that supports a set of interesting operations with good concurrency and performance while simultaneously allowing mutability.
We began this project with the following goals: 1)~enable efficient, concurrent mutable insertion and deletion operations at a cheaper cost than completely rebuilding the data structure; 2)~support a full set of queries (e.g., lookup, count and range); and 3)~allow parallel execution of all these operations to efficiently exploit the massive parallelism offered by modern GPUs while guaranteeing correctness at any time.

\section{Background and Previous Work}\label{sec:background}
\remove{
\paragraph{GPUs as General-Purpose Processors}
The modern GPU
is characterized by a large number of parallel compute units, often grouped in a SIMD configuration, and simple control hardware, typically controlling an entire SIMD unit in lockstep. It features a computation hierarchy visible to software, typically including individual compute units (``thread processors''), SIMD collections of compute units (``warps''), and multiple cores (NVIDIA's ``streaming multiprocessors'' or SMs) that can independently run blocks up to on the order of 1k threads.
We use CUDA~\cite{NVIDIA:2015:CUDA} as our programming language, which is built atop the C programming language and offers low-level access to hardware.
Programmers write code for a single thread in a SPMD fashion as a ``kernel''; the hardware is then responsible for scheduling blocks of threads to different SMs.

GPUs feature a memory hierarchy: on NVIDIA GPUs, which are typical, each thread has its own local registers; threads within a block can quickly share data through a fast, fixed-size, programmer-controlled ``shared memory''; SMs each have an independent L1 cache and all SMs share an L2 cache; and a high-bandwidth external DRAM memory holds global data accessible to all threads.
To avoid in-memory communications, NVIDIA GPUs also support fast warp-wide voting schemes (such as \code{ballot}) that we use in our implementation.
Because global DRAM bandwidth is often the bottleneck for GPU programs, it is crucial for performance to structure global memory accesses as much as possible such that neighboring threads access neighboring locations in GPU memory (``coalesced memory accesses'').
}
\paragraph{GPU Data Structures}
For general-purpose use, GPUs basically have three data structures: an unordered array, a sorted array, and a hash table. Application-specific data structures, such as acceleration tree data structures used in ray tracing~\cite{Zhou:2008:RKC}, may also be useful for general-purpose tasks.
Historically, complex data structures used by the GPU have been built on the CPU and then copied to the GPU \replace{(for example, perfect spatial hash tables~\cite{Lefebvre:2006:PSH}), but recent work in, for example, cuckoo~\cite{Alcantara:2009:RPH:nourl} and Robin Hood hashing~\cite{Garcia:2011:CPH} and bounding volume hierarchy (BVH) trees~\cite{Karras:2012:MPI} has focused on data structures that can be efficiently constructed directly on the GPU\@.}{but recent work in, for example, cuckoo hashing~\cite{Alcantara:2009:RPH:nourl} and bounding volume hierarchy (BVH) trees~\cite{Karras:2012:MPI} has focused on data structures that can be efficiently constructed directly on the GPU\@.}


There have been some efforts to enhance performance on range queries in particular data structures.
\remove{For example, Yang et al.\ implemented grid files (built on the CPU) for multidimensional database queries~\cite{Yang:2007:IGF}.
Kim et al.\ built an architecture sensitive binary tree~\cite{Kim:2010:FFA} for both CPUs and GPUs.}
Fix et al.\ implemented a brute force GPU method for lookup and range queries on a CPU-built B+ tree, targeted for database systems~\cite{Fix:2011:ABB}.
Range queries have also been used for spatial data structures such as R-trees~\cite{Luo:2012:PIR} or for processing moving objects~\cite{Liao:2014:ACR}.
In general,
however, there are not yet any such data structures that support general updates on the GPU\@.
All the work we cite above requires a complete rebuild for updating the data structure; and in fact
most of them can only be built on the CPU\@.

If range and/or counting queries were not of interest, we might choose hash tables~\cite{Alcantara:2009:RPH:nourl}
or non-clustered B-trees.
Prior GPU work in these areas is also not efficiently updatable, so this is also an interesting area of research, which we
do not address in this paper.


%


\paragraph{In-memory dictionaries}
Dozens of sequential data structures implement dictionaries,
including 2-3 trees, red-black trees, AVL tree, treaps, skip lists,
and many more~\cite{CormenLeRi09}.
These data structures are primarily oriented for use on
CPUs when the data is memory-resident; they are not optimized for locality
of reference.
They all share the trait that they have $O(\log n)$
levels, and each level might induce a memory fault.
\revision{Throughout this work, $n$ is the total number of elements in the data structure.}

\paragraph{External memory dictionaries}
A different, but also large, set of data structures were devised for cases where memory
faults dominate the cost of using the dictionary.
Because this setting emphasizes locality of reference, it has
interesting analogies to the GPU context.
The most widely used external memory dictionary
is the B-tree.
A B-tree is a
search tree in which each non-leaf node (except possibly the root) has
$\Theta(B)$ children, where $B$ is chosen so that $B/2$ keys fit in a
page \revision{(in the external memory model, memory accesses are assumed to be done in blocks of size $B$)}.
Since all the keys in a page are
fetched together, a B-tree is faster than a balanced binary tree by a
factor of $\log B$, which can be quite significant, depending on the
size of keys and the size of pages.

B-trees enjoy an optimal number of page misses during
searches, but they are suboptimal for insertions, deletions and
updates~\cite{BrodalFa03b}.
There are some data structures that target better insertions while providing optimal searches (or somewhat worse).
The two most influential data structures among these are the inspirations for this
work: the Log-structured Merge-tree (LSM)~\cite{Oneil:1996:LSM}
and the Cache Oblivious Lookahead Array (COLA)~\cite{Bender:2007:CSB}.
The basic idea of an LSM is to keep a set of dictionaries, each a
constant factor larger than the one before.
As insertions arrive,
they are placed in a small dictionary.
When a dictionary reaches
capacity, it is merged with the next larger dictionary, and so on.
In an LSM, a query must be performed at each level.  Therefore, LSMs
in the external memory context
usually have a B-tree at each level.  Since $O(\log n)$ searches must
be performed, the search time in an LSM takes $O(\log n \log_B n)$ I/Os in
the worst case, making them slower for queries than B-trees.  The
improvement comes from reducing the number of I/Os needed to insert.
Also, since items are moved from
one level of the LSM to the next in a batch, the insertions in the
next level enjoy some locality of reference.
LSMs are asymptotically much faster than straight B-trees for
insertions.

A COLA replaces the B-tree at each level with a sorted array.
This would seem to make searches slower, since, as noted above, B-trees
perform searches faster than binary search into an array.
To get fast search times,
pointers are maintained between levels via fractional
cascading.
Thus COLAs enjoy the same search speed as B-trees
while achieving the same insertion complexity as the LSM\@.

\section{The GPU LSM}\label{sec:gpu_cola}
In this paper, we propose a GPU dictionary data structure---the \emph{GPU LSM}---that combines the general structure of the LSM and the use of sorted arrays as in the COLA\@.
The usual LSM implementation, with a  B-tree per level, is not cache-oblivious,
which makes it a poor match for GPU hardware with relatively small cache sizes.\footnote{For example, the NVIDIA Tesla K40c provides 16~KB L1-cache per SM, 1.5~MB L2-cache to be shared among all SMs, and 48~KB manually managed shared memory per SM\@.}
We instead use a sorted array -- a cache-oblivious data structure -- at each level, like the COLA\@.
We do not, however, maintain the COLA's inter-level pointers for fractional cascading,
because of the implementation complexity and performance implications, especially for parallel updates.
Theoretically this means that we perform lookup
queries in $O(\log^2 n)$ compared to the optimal $O(\log n)$ achieved by the COLA~\cite{Bender:2007:CSB}).
In practice, we observe (Section~\ref{subsec:perf_lookup}) that this deficiency does not affect our query results much.

As we shall see, we can implement the
updates in a consistent way using only two primitives, sort and merge.
Both sort and merge are bulk primitives that have very efficient parallel formulations for GPUs.
As a result, we will see that we can handle high rates of insertions and deletions.
Having a sorted array at each level also gives us
good locality of reference for retrievals, a requirement for any high-performance GPU program.
Finally, the hierarchical, exponentially scaled, structure of the LSM and COLA
almost always prevents
significant parts of the data structure from being touched during insertions or
deletions.

\subsection{Batch Operation Semantics}\label{subsec:cola_design}
As discussed in Section~\ref{sec:intro},
we focus on \emph{batch} operations,
consistent with the GPU's bulk-synchronous programming model.
For queries this is straightforward: they do not modify the
data structure, so multiple queries can run
simultaneously without correctness issues.\remove{\footnote{For completeness, we have also considered individual query implementations, where, for instance, a single thread can issue a new lookup query asynchronously and independently from others.}}
For updates, however, correctness becomes challenging.
We begin by defining the specific semantics we use in batch operations:
\begin{enumerate}
\item The GPU LSM has a fixed batch size,
defined by a parameter $b$, which is also the size of the first level.
The choice of $b$ is application and platform dependent, and can help trade off query and update performance;
\item All update operations are in a batch of size $b$. We may have mixed batches of insertions and deletions.
Query operations can be in batches of any size.
Updates and queries are performed in separate phases;
\item If we insert items with the same key in different batches,
the most recently inserted is the only valid value of the key.
All previously inserted items with that key are termed \emph{stale}.
The notion of time is discretized based on the order of batch insertions;
\item If we insert multiple items with the same key in the same batch,
an arbitrary one is chosen as the current valid item;
\item After deleting a key, all previous instances of that key
are considered deleted and become stale.
Multiple deletions of the same key within a batch have the same effect as
one deletion; and
\item A key that is inserted and deleted within the same batch is considered
deleted.
\end{enumerate}
Our strategy for efficiently implementing the update semantics will be to
tolerate the presence of some stale elements in the data structure, so long as they
do not affect current query results.
Periodically, the user can choose to clean up (flush) the data structure, removing
stale elements and improving query efficiency.

\subsection{Insertion}\label{subsec:cola_insertion}
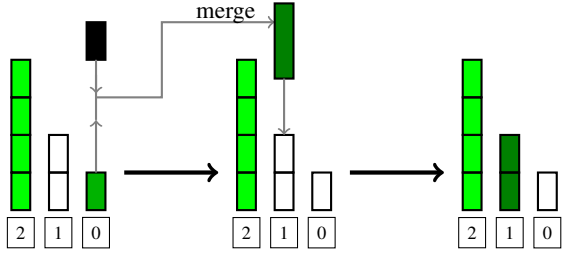
\begin{figure}
\centering
\begin{tikzpicture}
\def\recX{0.25}
\def\recY{0.5}
\def\idx{1}
\def\dx{0.5}
\foreach \x in {0,...,3}
	\fill [green] (0 * \dx,\x * \recY) rectangle (0 * \dx + \recX, \x * \recY + \recY);

\foreach \x in {0,...,3}
	\draw [black, thick] (0 * \dx,\x * \recY) rectangle (0 * \dx + \recX, \x * \recY + \recY);

\node[draw] at (0.125,-0.3) {\scriptsize 2};
\node[draw] at (0.625,-0.3) {\scriptsize 1};
\node[draw] at (1.125,-0.3) {\scriptsize 0};

\foreach \x in {0,1}
	\draw [black, thick] (\dx,\x * \recY) rectangle (\dx + \recX, \x * \recY + \recY);

\fill [black!30!green] (2 * \dx,0) rectangle (2 * \dx + \recX, 0  + \recY);
\draw [black, thick]	(2 * \dx,0) rectangle (2 * \dx + \recX, 0 + \recY);

\fill [black] (1,2.0) rectangle (1 + \recX, 2.0 + \recY);
\draw [black,thick] (1,2.0) rectangle (1 + \recX, 2.0 + \recY);

\draw [gray, thick, ->] (1.125, \recY) 	-- (1.125, \recY * 0.5 + 2.5 * 0.5 - 0.3);
\draw [gray, thick, ->] (1.125, 2.0) 	-- (1.125, \recY * 0.5 + 2.0 * 0.5 + 0.3);
\draw [gray, thick] 	(1.125, \recY/2 + 2.5/2 - 0.3) 	-- (1.125, \recY/2 + 2.5/2 + 0.3);

\draw [gray, thick, ->] (1.125,\recY/2 + 2.5/2) -- (2,\recY/2 + 2.5/2) -- (2,5*\recY) -- (3.5,5*\recY); 
\node[] at (2.85,5*\recY + 0.1) {\small merge};

\foreach \x in {0,...,3}
	\fill [green] (3 + 0 * \dx,\x * \recY) rectangle (3 + 0 * \dx + \recX, \x *\recY + \recY);

\foreach \x in {0,...,3}
	\draw [black, thick] (3 + 0 * \dx,\x * \recY) rectangle (3 + 0 * \dx + \recX, \x * \recY + \recY);

\foreach \x in {0,1}
	\draw [black, thick] (3 + \dx,\x * \recY) rectangle (3 + \dx + \recX, \x * \recY + \recY);

\draw [black, thick]	(3 + 2 * \dx,0) rectangle (3 + 2 * \dx + \recX, 0 + \recY);

\fill [black!50!green] (3 + \dx, 1.75) rectangle (3 + \dx + \recX, 1.75 + 2 * \recY);
\draw [black, thick] (3 + \dx, 1.75) rectangle (3 + \dx + \recX, 1.75 + 2 * \recY);

\draw [gray, thick, ->] (3 + \dx + \recX/2,1.75) -- (3 + \dx + \recX/2,2*\recY);

\node[draw] at (3 + 0.125,-0.3) {\scriptsize 2};
\node[draw] at (3 + 0.625,-0.3) {\scriptsize 1};
\node[draw] at (3 + 1.125,-0.3) {\scriptsize 0};

\foreach \x in {0,...,3}
	\fill [green] (6 + 0 * \dx,\x * \recY) rectangle (6 + 0 * \dx + \recX, \x * \recY + \recY);

\foreach \x in {0,...,3}
	\draw [black, thick] (6 + 0 * \dx,\x * \recY) rectangle (6 + 0 * \dx + \recX, \x * \recY + \recY);

\foreach \x in {0,1}
	\fill [black!50!green] (6 + \dx,\x * \recY) rectangle (6 + \dx + \recX, \x * \recY+ \recY);
\foreach \x in {0,1}
	\draw [black, thick] (6 + \dx,\x * \recY) rectangle (6 + \dx + \recX, \x * \recY+ \recY);

\draw [black, thick]	(6 + 2 * \dx,0) rectangle (6 + 2 * \dx + \recX, 0 + \recY);

\node[draw] at (6 + 0.125,-0.3) {\scriptsize 2};
\node[draw] at (6 + 0.625,-0.3) {\scriptsize 1};
\node[draw] at (6 + 1.125,-0.3) {\scriptsize 0};

\draw [ultra thick, ->] (1.5, 0.5) -- (2.75, 0.5);
\draw [ultra thick, ->] (4.5, 0.5) -- (5.75, 0.5);
\end{tikzpicture}
\caption{\small Insertion example in GPU LSM; adding a new batch of $b$ elements into a GPU LSM with $5b$ elements. Blocks with similar colors are sorted among themselves.}
\label{fig:eg_insertion_cola}
\end{figure}

Insertions are where LSMs and COLAs really shine; in the
external memory context, their insertion performance far outperforms simple B-trees.
In the GPU LSM, since all insertions are in units of $b$ elements,
the size of level $i$ in the GPU LSM is $b2^i$, and at any time the whole data structure contains a multiple of $b$ elements.
Each level is completely full or completely empty.
To insert a  batch of size $b$, the batch is first sorted.
If the first level is empty, the batch becomes the new first level.
Otherwise, we merge the sorted batch with
the sorted list already residing in the first level,
forming a sorted array of size $2b$, and proceed to the second level, and so on.
%
Figure~\ref{fig:eg_insertion_cola} schematically shows the insertion process in the GPU LSM\@.
\revision{Figure~\ref{fig:alg_insertion} shows the high level structure of the insertion algorithm.}

In a GPU LSM with $n = rb$ elements (\emph{resident elements}),
the levels that contain sorted lists correspond to the set bits in the
the binary representation of the integer $r$.
The insertion process corresponds to incrementing $r$,
with additions and carries of binary arithmetic corresponding to the
merging procedure described above.
The larger our choice of $b$, the more parallelism can be exploited.
Smaller $b$ sizes lead to inefficiency in the first few levels for each update operation.

Note that insertion proceeds from smaller, lower indexed, levels to larger.
Thus any item at level $r$ must have been inserted before any element in a lower-indexed level.
We  make sure that after merging, insertion order is preserved among elements with the same key at the same level (Section~\ref{subsec:insertion_details}).

The original analyses of LSMs and COLAs were based on memory block size (see Bender et al.~\cite{Bender:2007:CSB}).
We apply the same observations to batches to show that any sequence of $r$ batch insertions
requires at most
$O(rb \log r)$ work, that is, $O(\log r)$ work per item.
Certainly a worst-case individual insertion requires a cascade of merges, ultimately
placing the final list in the level corresponding
to the most significant bit of $r$, and hence
time $\Omega(rb)$.
But the key idea is that such a worst-case operation can only occur infrequently.
In particular, an element residing in
a list of length $O(2^i b)$ has participated in $O(i) \leq O(\log r)$
merge operations, so that the total work performed
over all $rb$ elements residing in the GPU LSM at any time must be $O(rb \log r)$.
\subsection{Deletion}\label{subsec:cola_deletion}
The standard way to delete an item in a
COLA is to insert a \textit{tombstone} item with the same key, indicating that previously inserted items with that key,
if any, should be considered deleted.
Deletion, then, is the insertion of a tombstone, making
deletion a kind of insertion \revision{(Figure~\ref{fig:alg_insertion})}, so that
we can combine any insertion and deletion requests into a mixed batch.
As we shall see in the following sections, this tombstoning scheme allows the
GPU LSM to perform insertions and deletions very efficiently, at the
cost of accumulating stale elements.
\subsection{Lookup}\label{subsec:cola_lookup}
Recall that \textsc{lookup}$(k)$ should
return the most recently inserted value corresponding to $k$, if it was
not subsequently deleted, or otherwise report that such a key does not exist.
To ensure this, we guarantee the following \emph{building invariants}
during insertion and deletion:
\begin{enumerate}
\item Within each level, all elements are sorted by key and
thus all elements with the same key are next to each other, forming a segment;
\item \revision{All elements within each segment
(regular elements and tombstone)
are ordered from the lowest index to the highest based on the
time they were inserted, from most recent to least recent;}
\item \revision{Tombstones within a segment
are placed before regular elements with the same key.}

\end{enumerate}
With these invariants,
it suffices to start our lookup process from the
smallest (most recent) occupied level and look for the
smallest index with key greater than or equal to $k$.
If we find a regular element with key equal to $k$, we return it and are done.
If we find a tombstone with key $k$,
then $k$ is deleted and we return no result (indicating that $k$ was not found).
Otherwise we found no element with key $k$, and we continue to the next occupied level.
\revision{A high level description of a set of lookup operations is shown in Fig.~\ref{fig:alg_lookup}.}

With $n = rb$ total elements, finding the lower bound (a modified
binary search) in each level takes $O(\log(b2^i))$ steps over $\log r\!$
steps, which in the worst case results in $O(\log^2(r) + \log (r)\log (b))$ individual memory
accesses per query (the same cost as in the basic LSM).
\subsection{Count and Range queries}\label{subsec:cola_range}
Both \textsc{count} and \textsc{range} operations take a tuple $(k_1,k_2)$ as an input argument. The former returns the total number of elements within that range of keys while the latter returns all those valid elements as an output. \revision{\textsc{count} and \textsc{range} are shown in Fig.~\ref{fig:alg_count} and \ref{fig:alg_range} respectively.}
Based on the same building semantics described in Section~\ref{subsec:cola_lookup}, we implement the following procedure.

Since each full level is sorted by key, we can use binary search to find indices corresponding to the first key $k$ such that $k \geq k_1$ or $k > k_2$ (also known as lower bound and upper bound operation, respectively).

If duplicates and deletions were not allowed in the data structure, we could compute a result by subtracting these indices to have the number of elements within those bounds (for \textsc{count}), or just collecting all elements within those bounds (for \textsc{range}).
However, with duplicates and deletions allowed, this approach would include three types of extra elements in our results: 1) tombstones, 2) deleted elements, and 3) elements with the same key that were replaced by later insertions.
To return an accurate answer, we perform a post-processing stage (discussed in Sections~\ref{subsec:count_details}--\ref{subsec:range_details}) to correct our preliminary potential results
(Afshani et al. discuss why counts are harder than lookups~\cite{afshani:2017:CRD}).
It should be noted this extra validation could be avoided if we were not allowing stale elements in the data structure (item 3 in Section~\ref{subsec:cola_design}), but that is the price that we pay to support deletion.
\begin{figure}
\centering
\algsetup{linenosize=\ssmall}
\subfloat[][\revision{\textsc{Insert} and \textsc{Delete}}]{
\begin{minipage}[t][3.7cm][c]{0.5\linewidth}
\ssmall
        \begin{algorithmic}[1]
        \STATE \textbf{Insert(batch) \{}
        \STATE \textbf{Input}: a batch of $b$ elements
        \STATE buffer $\leftarrow$ sort(input)
        \STATE i $\leftarrow$ 0
        \WHILE{level i is full}
        \STATE buffer $\leftarrow$ merge(buffer, level i)
        \STATE level i $\leftarrow$ 0
        \STATE i++
        \ENDWHILE
        \STATE level i $\leftarrow$ buffer
        \RETURN
        \STATE \}
        \STATE \textbf{Delete(batch) \{}
        \STATE \textbf{Input}: a batch of $b$ elements
        \STATE Insert(tombed(input))
        \RETURN
        \STATE \}
        \end{algorithmic}
\end{minipage}\label{fig:alg_insertion}
}
\subfloat[][\revision{\textsc{Lookup}}]{
\ssmall
\begin{minipage}[t][3.7cm][c]{0.5\linewidth}
        \begin{algorithmic}[1]
        \STATE \textbf{Lookup(k)} \{
        \STATE \textbf{Input}: a set of lookup queries
        \FOR{each query k \textbf{in parallel}}
        \FOR{i = 0 to the last level}
        \IF{level i is full}
        \STATE result $\leftarrow$ lower\_bound(level i, k)
        \IF{result is valid}
        \RETURN result
        \ENDIF
        \ENDIF
        \ENDFOR
        \RETURN $\perp$
        \ENDFOR
        \STATE \}
        \end{algorithmic}
\end{minipage}\label{fig:alg_lookup}
}\\
\subfloat[][\revision{\textsc{Count}}]{
\ssmall
\begin{minipage}[c]{0.5\linewidth}
        \begin{algorithmic}[1]
        \STATE \textbf{Count(k1,k2) \{}
        \STATE \textbf{Input}: a set of count queries
        \FOR{each query q = (k1, k2) \textbf{in parallel}}
        \FOR{i = 0 to the last level}
                \STATE l[q][i] $\leftarrow$ lower\_bound(level i, k1)
                \STATE u[q][i] $\leftarrow$ upper\_bound(level i, k2)
                \STATE init\_count[q][i] $\leftarrow$ u[q][i] - l[q][i] + 1
        \ENDFOR
        \ENDFOR
        \STATE offset $\leftarrow$ exclusive\_scan(init\_count)
        \FOR{each query q = (k1, k2) \textbf{in parallel}}
        \FOR{i = 0 to the last level}
                \STATE storing all elements in level i from l[q][i] until u[q][i] into result[q][offset[q][i]]
        \ENDFOR
                \STATE // marking all invalid (stale) elements
                \STATE result[q][:] $\leftarrow$ post\_process(result[q][:])
                \STATE count[q] $\leftarrow $ No.\ of valid elements in result[q][:]
                \RETURN count[q]
        \ENDFOR
        \STATE \}

        \end{algorithmic}
\end{minipage}\label{fig:alg_count}
}
\subfloat[][\revision{\textsc{Range}}]{
\ssmall
\begin{minipage}[c]{0.5\linewidth}
        \begin{algorithmic}[1]
        \STATE \textbf{Range(k1,k2) \{}
        \STATE \textbf{Input}: a set of range queries
        \FOR{each query q = (k1, k2) \textbf{in parallel}}
        \FOR{i = 0 to the last level}
                \STATE l[q][i] $\leftarrow$ lower\_bound(level i, k1)
                \STATE u[q][i] $\leftarrow$ upper\_bound(level i, k2)
                \STATE init\_count[q][i] $\leftarrow$ u[q][i] - l[q][i] + 1
        \ENDFOR
        \ENDFOR
        \STATE offset $\leftarrow$ exclusive\_scan(init\_count)
        \FOR{each query q = (k1, k2) \textbf{in parallel}}
        \FOR{i = 0 to the last level}
                \STATE storing all elements in level i from l[q][i] until u[q][i] into result[q][offset[q][i]]
        \ENDFOR
                \STATE // removing all stale elements
                \STATE result[q][:] $\leftarrow$ post\_process(result[q][:])
                \STATE result[q][:] $\leftarrow$ compact(result[q][:])
                \STATE count[q] $\leftarrow $ No.\ of valid elements in result[q][:]
                \RETURN result[q][0:count[q]-1]
        \ENDFOR
        \STATE \}
        \end{algorithmic}
\end{minipage}\label{fig:alg_range}
}
\caption{\small \revision{Pseudocode for the GPU LSM's operations. Insertion and deletion  operate on a fixed-size batch of size $b$. Lookup, count and range are processed in parallel on a set of input queries. In order to simplfy the pseudocode we have only considered keys (no values).}}\label{fig:alg_overview}
\end{figure}

\subsection{Cleanup}\label{subsec:cola_cleanup}
The structure we have described so far produces correct query results
even in the presence of tombstones and stale elements (either deleted or
duplicate).
This allows us to perform faster insertions and deletions,
but as tombstones and stale elements accumulate,
we will see increased memory usage and more occupied levels,
resulting in reduced query performance.
Thus we provide a \textsc{cleanup} operation, in which all tombstones,
their corresponding deleted elements, and all duplicates (i.e., replaced elements) are removed, followed by a reorganization of the GPU LSM\@.

In the theoretical sense, periodic cleanups can be done without increasing the total work.
We see this by noting that if a cleanup is performed after $O(rb)$ operations, its cost is at most
$O(rb)$:
rebuilding the GPU LSM from scratch requires a radix sort, compaction
to remove stale elements and tombstones, and slicing up the remaining sorted
list of valid elements into a sequence of levels.
A schedule in which a cleanup is performed every time the GPU LSM doubles in size
can thus be amortized so that the asymptotic total work does not change; that is, the
work done in cleanup can be ``charged" to the sequence of operations that triggers it.

When is a cleanup appropriate in practice? Cleanups are not free, but they may reduce the
number of LSM levels and increase query performance when there have been
a significant number of deletions and/or replacements.
Frequent cleanups also may be appropriate in applications where query performance is paramount.

\section{Implementation details}\label{sec:impl_details}
In this section we dive deeper into the implementation details of each
operation and the design choices that we make.
Our design decisions are partially influenced by hardware characteristics (NVIDIA GPUs), programming environment (CUDA), and available open-source GPU primitive libraries (moderngpu\footnote{Moderngpu is available at \url{https://github.com/moderngpu/moderngpu}.} and CUB~\cite{Merrill:2015:CUB}).
Our GPU LSM is built using 32-bit variables (both keys and values).
We use the term ``element'' to refer to what is stored at a
specific index in arrays
(key arrays, value arrays, etc.) allocated in the GPU's global memory.
\revision{In this work, all major operations (including sort and merge) are done on the GPU\@.}
\subsection{Insertion and Deletion}
\label{subsec:insertion_details}\label{subsec:deletion_details}
In Sections~\ref{subsec:cola_insertion}~and~\ref{subsec:cola_deletion}, we
noted that the implementation of insertion and deletion in the GPU LSM
were similar.
In order to distinguish between a tombstone and a regular element
to be inserted,
we dedicate one bit
as a flag; we refer to this bit as the \emph{status bit}.
The 32-bit \emph{key variable} is the 31-bit \emph{original key} shifted once and placed next to the status bit.
The cost of this decision is that we lose one bit in the key domain.

\revision{We outline our insertion procedure in Fig.~\ref{fig:insert_detail}, where for the sake of clarity we assume a key-only (no values) scenario.}
As described in Section~\ref{subsec:cola_insertion}, when
inserting a batch of $b$ elements, we first sort the new batch
by their keys.
In order to fulfill the building invariants of
Section~\ref{subsec:cola_lookup} while also satisfying
the semantics
in Section~\ref{subsec:cola_design}, we implement sorting and merging as follows.
For sorting, we use regular radix sort over all key variables including the status bit \revision{(line 9 in Fig.~\ref{fig:insert_detail})}.
For merging, we merge different levels just based on the original keys, excluding the status bit \revision{(line 14 in Fig.~\ref{fig:insert_detail})}.
We make sure that stability is preserved: new levels merged into existing levels appear first in the merged result, to preserve batch insertion order.
We use moderngpu's merges with a modified comparison operator and CUB's radix sort for this process.
\revision{Since our merge is not an in-place operation, we use double buffers and a ping-pong strategy between them (not shown in Fig.~\ref{fig:insert_detail}).}


Since a tombstone is marked with a zero LSB,
after sorting a batch prior to insertion, a tombstone will appear before
any insertions with the same key.
And, since our merge is stable and we neglect the status bits,
a regular element from a smaller (more recent) level always appears
before all later elements with the same key (regular or tombstone)
that came from previously inserted batches.
Note that after merging we do not remove stale elements.
Nonetheless, because those stale elements appear after the
replacement or tombstone
element that made them stale, they are ``invisible''
to the queries.
In case there are only $b^\prime < b$ new elements to be inserted, a user can pad the batch by duplicating enough ($b - b^\prime$)
copies of an arbitrary element within the batch (e.g., the last one);
only one of those duplicates will be visible to queries.
If our GPU LSM achieves an average $R$ insertion rate, here we are effectively only using $Rb^\prime /b$ of our available insertion rate.
As a result, users should choose $b$ wisely to balance the cost of partial batches with the efficiency of larger batches.

\revision{\paragraph{Cache and shared memory usage}
In the original LSM data structures on the CPU, higher levels of the data structure are small enough to fit in cache, so insertions are effectively performed inside cache before getting merged with lower levels of the data structure.
In our GPU LSM, based on our choice of having a sorted array (a cache-oblivious data structure) in each level of the LSM, higher levels will automatically cache efficiently, aided by the high associativity of the L2 cache on the GPU\@.
Nevertheless, insertions/deletions are only done in batches of size $b \gg 1$, which means fewer levels can be fit into cache, but each level has more elements in it (making it just as practically effective as CPU implementations).
As well, the sort and merge primitives we chose aggressively use shared memory to achieve coalesced global memory accesses so that final results (sorted/merged chunks of input) are first stored into shared memory and then stored into global memory in larger batches.}
\begin{figure}
\centering
\ssmall
\algsetup{indent=2em}
\algsetup{linenosize=\ssmall}
\begin{algorithmic}[1]
\STATE {class GPU\_LSM\{}
\STATE {uint \qquad \qquad b; \qquad \qquad \qquad // batch size}
\STATE {uint \qquad \qquad num\_batch; \qquad // number of inserted batches}
\STATE {KeyType* \quad \enskip  d\_lsm\_key;} \qquad // an array of keys for all levels
\STATE {//=======}
\STATE // Input: a mixed batch of regular insertions and tombed deletions
\STATE {{\textbf{Insert(KeyType \quad keys[0:b-1])}\{} }
\STATE {// sorting including the LSB, and storing the result into d\_buffer}
\STATE {d\_buffer $\leftarrow$ CUB::sort(keys, [](KeyType x, KeyType y)\{return x $<$ y;\});}
\STATE {i = 0; \qquad offset = 0;}
\STATE {n\_level\_i = b; \quad // number of elements in level i}
\WHILE{((num\_batch \& (1 $\ll$ i)) != 0)}
\STATE {// merging excluding the LSB}
\STATE {d\_buffer $\leftarrow$ mgpu::merge(d\_lsm\_key + offset, n\_level\_i, d\_buffer, n\_level\_i, [](KeyType x, KeyType y)\{return (x $\gg$ 1) $<$ (y $\gg$ 1)\});}
\STATE {gpu\_memory\_set(d\_lsm\_key + offset, 0, n\_level\_i); // emptying the previous level}
\STATE {i++;}
\STATE {offset += n\_level\_i; \quad n\_level\_i $\ll$= 1;}
\ENDWHILE
\STATE {// Inserting into the first empty level:}
\STATE gpu\_memory\_copy(d\_lsm\_key + offset, d\_buffer);
\STATE {num\_batch++;}
\STATE {\textbf{\}}}
\STATE \};
\end{algorithmic}
\caption{\small \revision{Implementation details for the Insert operation, which inserts an arbitrary mixed batch of new elements to be inserted and tombed elements to be removed.} \label{fig:insert_detail}}
\end{figure}

\subsection{Lookup queries}\label{subsec:lookup_details}
We could implement lookup queries in two ways: a bulk approach and an individual approach.
For the bulk approach, we would first sort all queries and then perform a sorted search (similar to moderngpu's) over all queries and all occupied levels.
However, sorting all queries is an expensive operation and having all queries ready synchronously may also be a strong assumption.
Thus we focus on an individual approach where each query can be performed independently.

Each thread can individually perform a lookup query by simply performing a lower-bound binary search
in each occupied level, starting from the smallest level.
There are three possible outcomes: 1) we find a regular element (set LSB) that matches our query and then return its value; 2) we find a tombstone (zero LSB) that matches our query and then return $\perp$; or 3) we search in all levels and nothing is found (return $\perp$).
The main bottleneck for our lookups is the random memory accesses required in all binary searches that we perform.
\subsection{Count queries}\label{subsec:count_details}
Count queries provide opportunities for concurrency and collaboration between queries.
We assume there are multiple queries to be performed (each with its $(k_1,k_2)$ argument) and each query can have any arbitrary size as its outcome.
We assign each GPU thread to a query and then perform device-wide operations to find all results together (with possible collaboration among threads).
The procedure takes five steps:
(1)~Initial count estimate:
we perform two separate binary searches for the
lower and upper limits of each query for each level.
By subtracting resulting indices at each level,
we get an upper bound on
the number of potentially valid elements \revision{(lines 4--8 in Fig.~\ref{fig:alg_count})}.
For each query, we write the
per-level upper bounds next to each other in memory.
(2)~Scanning stage:
we perform a device-wide scan operation over the
count estimates to compute
a global index for each query result \revision{(line 10 in Fig.~\ref{fig:alg_count})}.
(3)~Initial key storage: we use
the lower limits of the queries and the
global indices computed in the previous step
to copy all potential keys into an array \revision{(lines 12--14 in Fig.~\ref{fig:alg_count})}.
(4)~Segmented sort: we perform a segmented sort over
the array (each segment belongs to a query).
LSBs (status bits) are neglected in sorting comparisons.
\revision{This stage is shown under the post-processing procedure on line 16 in Fig.~\ref{fig:alg_count}.}
(5)~Final counting \revision{(line 17 in Fig.~\ref{fig:alg_count})}: all identical keys (tombstones or regular)
are now next to each other and sorted by time step.
We count the first element of each segment only if it is not a
tombstone. As a result, if there are multiple keys, only one of them is counted.

We use CUB's exclusive scan (prefix-sum) for stage 2.
For stage 3, we assign each thread to a query but force the threads in a warp to collaborate with each other in writing out results from all 32 queries involved (for coalesced accesses).
We use moderngpu's segmented sort for stage 4.
In stage 5, we assign each query to a thread but force the threads in a warp to collaborate with each other in validating and counting (via warp-wide ballots) the results for all potential matches from 32 consecutive queries.

\subsection{Range queries}\label{subsec:range_details}
We implement \textsc{range} operations similarly to \textsc{count},
differing in the final three steps \revision{(lines 17--19 in Fig.~\ref{fig:alg_range})}.
In stage 3, we store not only all potential keys from each query but also their corresponding values.
Stage 4 becomes a segmented sort over key-value pairs rather than just keys.
Finally, in stage 5, we first mark each valid element (by overwriting the LSBs)
and then perform a segmented compaction based on all set LSBs (each segment represents a query) to gather all non-stale non-tombstone
elements.
The final result is the beginning memory offsets of each query, followed by valid elements (both keys and values) belonging to each query, sorted by their keys.
\subsection{Cleanup}\label{subsec:cleanup_details}
As we described in Section~\ref{subsec:cola_cleanup}, \textsc{cleanup} removes all stale elements (tombstones, deleted and duplicates) and rebuilds the data structure.
In its implementation, we require 1) cross-validation among different levels to remove deleted and duplicate elements from different levels (caused by elements from more recent levels) and 2) that the total number of elements should remain a multiple of $b$ after removals.
To satisfy the first item, we make sure all our intermediate operations within the cleanup are ``stable'', meaning that among all instances of an arbitrary key (tombstone or regular), temporal information is preserved.
The second item is important since we assume
that the number of elements in our data structure is
a multiple of $b$, and this may no longer be true
after removing stale elements.
We could either remove enough tombstones and duplicates to
make sure the condition is satisfied and leave the rest unchanged
(to be processed in future cleanups),
or we can simply remove all stale elements and then pad
with enough ($<b$) \emph{placebo} elements.\footnote{We pad our data structure with tombstone status bits and maximum keys (e.g., $2^{32}-1$ in 32-bit scenarios) at the end of the last level's sorted array. These elements will
be invisible to queries since they are tombstones, and will always remain at the end of the last level since no larger key is possible.}
We choose the second approach because it requires less processing and appears to be more efficient.

Our bulk strategy for cleanup is to 1)~iteratively merge all occupied levels from the smallest to the largest (neglecting the LSB to preserve time ordering), 2)~mark all unmarked stale elements \revision{(e.g., overwriting the LSBs)}, 3)~compact all valid elements together \revision{(e.g., using a two-bucket multisplit~\cite{Ashkiani:2016:GM:nourl} to collect all unmarked valid elements in stage~2)}, 4)~add enough placebos, and 5)~redistribute (already sorted) elements to different (new) levels.
Since all levels are already sorted, merging them together iteratively is much faster than resorting all of them together.
Our cleanup implementation preserves timing order among elements with the same original key, but not across different keys (smaller keys will end up in smaller levels).

\section{Performance Evaluation}\label{sec:perf_eval}
\subsection{What are the appropriate comparisons?}\label{subsec:perf_overview}
We compare our results with two data structures (Table~\ref{table:perf_comparison}): a GPU hash table (cuckoo hashing~\cite{Alcantara:2009:RPH:nourl}), and a \revision{GPU-maintained} sorted array (GPU SA\revision{, implemented by the authors}).
As noted in the introduction, neither of these data structures are mutable.
Cuckoo hashing\footnote{Cuckoo hashing code is from CUDPP: \url{https://github.com/cudpp/cudpp}.} 
has bulk build and lookup operations, but it does not support deletions 
and it is not possible to increase table sizes at runtime so insertion is possible only in a very 
limited sense. 
In the GPU SA, insertions (or deletions) can happen by adding (or removing) elements and resorting the whole array, which, as we shall see, is
much slower than applying updates to a GPU LSM\@.
Merging an already-sorted set of elements into an existing GPU SA, however, is faster than applying a set of sorted updates
to a GPU LSM\@.
All queries in a GPU SA are similar to those on the GPU LSM, but only on a single occupied level (of arbitrary size).
In general, all GPU SA queries (lookup, count, and range) have a faster worst-case scenario than the GPU LSM's, because it is easier to search a single sorted level ($O(\log n)$) than to search through multiple smaller occupied levels in the GPU LSM ($O(\log^2 n)$); in practice, though, we find that the difference is surprisingly small.

\begin{table}
\centering
\resizebox{\columnwidth}{!}{%
\begin{tabular}{c ccc}
\toprule
& Cuckoo hashing & Sorted Array (GPU SA) & GPU LSM \\
\midrule
\textsc{insert} & --- & $O(n)$ & $O(\log n)$ \\
\textsc{delete} & --- & $O(n)$ & $O(\log n)$ \\
\textsc{lookup} & $O(1)$ & $O(\log n)$ & $O(\log^2 n)$ \\
\textsc{count/range} & $O(n)$ & $O(\log n + L)$ & $O(\log^2 n+L)$ \\
\bottomrule
\end{tabular}%
}
\caption{\small High-level comparison of capabilities in a GPU hash table, a sorted array and a GPU LSM\@. Bounds on the total work are normalized to be
per item in a data structure with $n$ elements.
$L$ denotes the size of the output.}\label{table:perf_comparison}
\end{table}


Sorted arrays and hash tables are the only fully developed GPU data structures for general-purpose tasks with well-known and optimized performance characteristics (in contrast to other dictionary 
data structures like B-trees that are not yet fully developed on GPUs).
As a result, although neither of them is a mutable data structure, we believe comparing our GPU LSM against them is the right way to benchmark our data structure. 
This comparison is not meant to disprove the mutability capability of our GPU LSM but instead to expose the price we paid for achieving it (in terms of query performance).

As we noted in Section~\ref{sec:background}, some GPU dictionary
data structures are designed to support high-performance queries (although without any support for mutability), but
most of these are constructed sequentially on the CPU and then transferred into the GPU, 
because efficient parallel construction is itself challenging enough to be avoided.
B-trees, for example, would be an excellent point of comparison because they are 
more suitable for coalesced memory accesses on GPUs (compared to a binary search that requires random memory accesses).
However, to the best of our knowledge to date, 
there is not any publicly available B-tree library in which the structure is built on the 
GPU\@.\remove{\footnote{As an example, Fix et al.\ experimented on lookup queries on a CPU-built B+ tree~\cite{Fix:2011:ABB}. Their data structure is first built on the CPU and put into a contiguous block of memory so that it can be transferred easily to the GPU\@.
They assigned all threads within a thread-block to perform each single lookup query on their data structure.
On an NVIDIA GTX 480 and with 100k elements in their B+ tree, without considering memory transfer time between the CPU and the GPU, they reported 20~$\mu$s for each query on the GPU and 2~$\mu$s for performing the same query on their CPU (Intel Core 2 Quad Q9400).
We do not believe their implementation represents a competitive data structure from the B-tree family for comparison with our LSM\@. As we will see in Section~\ref{subsec:perf_lookup}, with 16~M elements, our GPU LSM can achieve an average lookup rate up to 75.7~M queries/s over 16~M queries (0.013~$\mu$s per query).}}


In the following sections
we discuss insertions and deletions in Section~\ref{subsec:perf_insertion}, then lookups, count, and range queries in Section~\ref{subsec:perf_queries}.
All experiments are run on an NVIDIA Tesla K40c GPU (ECC enabled), with Kepler architecture and 12~GB DRAM, and an Intel Xeon CPU {E5630}. All programs are compiled with NVIDIA's nvcc compiler (version~7.5.17) with the -O3 optimization flag.

\subsection{Insertions and deletions}\label{subsec:perf_insertion}
For any batch, including an arbitrary set of insertions and deletions, the GPU LSM provides the same performance (i.e., performance does not depend on status bits).
As a result, here we just consider pure insertion.
Our principal target is an application scenario where we repeatedly insert batches of elements into our data structure. On this scenario---which we quantify as ``effective insertion rate''---the GPU LSM has considerably better theoretical and experimental behavior than a hash table or a sorted array, as we will see at the end of this subsection. However, before we explore this scenario, we must first consider two preliminary scenarios: 1)~building the data structure from scratch (``bulk build'') and 2)~inserting one single batch into an already-built data structure (``batch insertion'').

\begin{table}
\centering
\resizebox{\columnwidth}{!}{%
\begin{tabular}{cccc || ccc}
\toprule
& \multicolumn{3}{c}{GPU LSM} & \multicolumn{3}{c}{Sorted Array (GPU SA)} \\
\cmidrule(lr){2-4}\cmidrule(lr){5-7}
$b$ & min rate & max rate & mean rate & min rate & max rate & mean rate \\

\midrule
$2^{27}$ & 727.8 & 727.8 & 727.8 & 727.8 & 727.8 & 727.8 \\
$2^{26}$ & 585.3 & 727.7 & 648.8 & 583.1 & 727.7 & 647.4 \\
$2^{25}$ & 421.1 & 727.1 & 585.5 & 472.2 & 726.8 & 561.2 \\
$2^{24}$ & 270.3 & 727.1 & 537.9 & 346.0 & 727.0 & 459.2 \\
$2^{23}$ & 155.5 & 726.3 & 485.2 & 224.0 & 723.2 & 334.0 \\
$2^{22}$ & 84.3 & 714.8 & 441.2 & 132.6 & 709.9 & 219.2 \\
$2^{21}$ & 44.0 & 694.0 & 398.1 & 74.2 & 691.2 & 131.2 \\
$2^{20}$ & 22.4 & 664.2 & 354.1 & 39.8 & 658.7 & 73.8 \\
$2^{19}$ & 11.3 & 558.2 & 289.4 & 20.5 & 555.9 & 39.4 \\
$2^{18}$ & 5.6 & 432.0 & 220.7 & 10.5 & 422.1 & 20.3 \\
$2^{17}$ & 2.8 & 326.8 & 159.9 & 2.6 & 319.1 & 10.3 \\
$2^{16}$ & 1.4 & 194.3 & 98.0 & 1.3 & 184.8 & 5.2 \\
$2^{15}$ & 0.7 & 101.4 & 58.4 & 0.6 & 95.9 & 2.6 \\
\midrule
mean & & & 225.3 &  &  & 16.7 \\
\midrule
\midrule
\multicolumn{2}{l}{Cuckoo Hash} & \multicolumn{5}{c}{361.7} \\
\bottomrule
\end{tabular}%
}

\caption{\small Minimum/maximum/harmonic mean insertion rates (M elements/s) for GPU LSM (and GPU SA) and with various batch sizes.
These numbers are recorded over all possible number of resident batches for $1 \leq r \leq 2^{27}/b$.
We also note the bulk build rate of cuckoo hashing with an 80\% load factor.}
\label{table:insertion_rate}
\end{table}

\paragraph{Bulk build} Suppose there are initially a set of $kb$ available elements from which we want to build a GPU LSM\@. This operation requires a sort and is similar to building a sorted array (GPU SA); the bulk build of either is faster than cuckoo hashing (up to 2x). Our GPU sustains 770 M~elements/s for key-value radix sort. Then we must segment this array into at most $\log k$ sorted levels corresponding to its GPU LSM levels (in-memory transfers with 288~GB/s = 36 G~elements/s); this time is negligible compared to sorting.\footnote{If we have multiple batches to insert into a non-empty GPU LSM, all batches can be sorted together and then iteratively merged from the highest valued batch until the lowest valued batch (to make sure deletions and time information are correctly preserved).}
Cuckoo hashing with an 80\% load factor has a build rate of 361.7 M~elements/s, roughly 2x slower than both the GPU LSM and the GPU SA\@.

\begin{figure}
\centering
\subfloat[][Batch insertion time (ms) in GPU LSM with $b = 2^{19}$.]{\includegraphics[width = 0.9\linewidth]{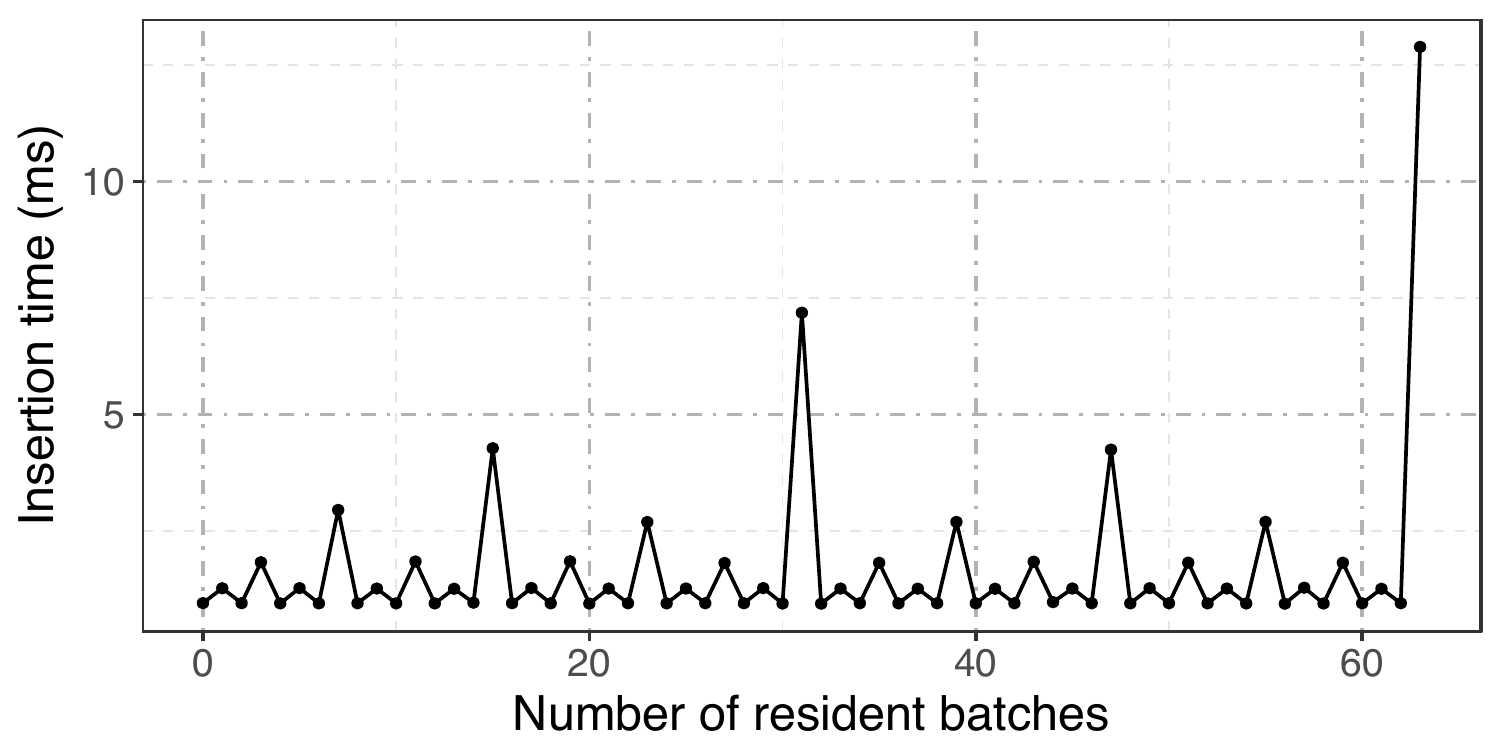}\label{fig:insertion_last_batch}} \\
\subfloat[][Effective insertion rate versus total number of inserted elements.]{\includegraphics[width = 0.9\linewidth]{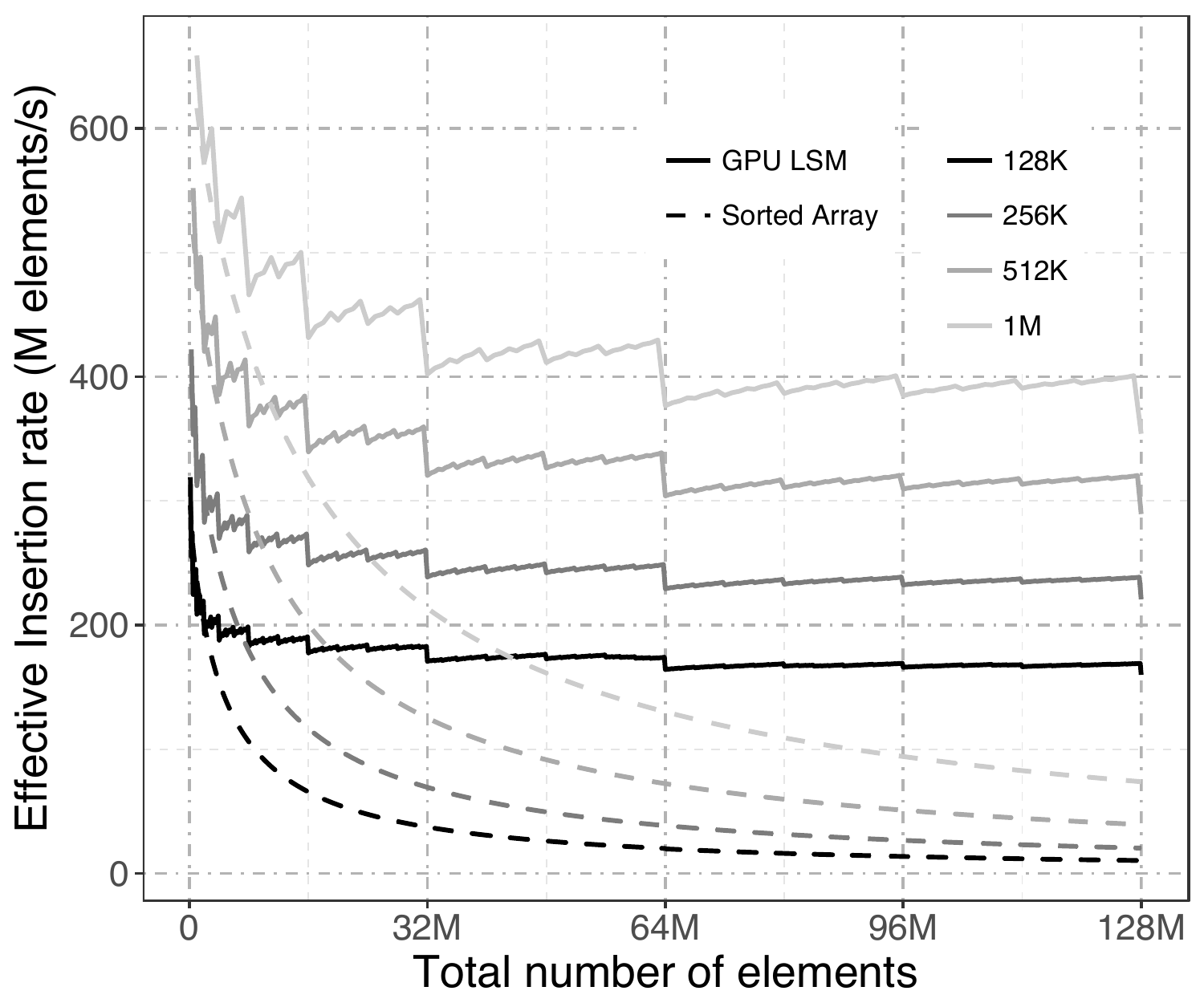}\label{fig:rate_insertion}} \\
\caption{(a) Batch insertion time in GPU LSM\@.
(b) Effective insertion rate for a series of batch insertions in GPU LSM, and a sorted array (GPU SA).
Each color represents a different batch size.}\label{fig:insertion}
\end{figure}

\paragraph{Batch insertion} To insert a new batch into an existing GPU LSM, we first sort the batch then iteratively merge with the lowest-indexed full levels until we reach the first empty level.
As a result, if there are $r$ resident batches in the GPU LSM prior to the new insertion, the first empty level's index will be the least significant zero bit in $r$ (we call it $\defn{ffz}(r)$).
The total amount of time required for inserting a new batch to a GPU LSM (total of $rb$ elements), also depicted in Fig.~\ref{fig:insertion_last_batch}, 
is
$
T_\text{ins}^{b}(r) = T_\text{sort}^{b} + (2^{\defn{ffz}(r)}-1) T_\text{merge}^{b},
$
where $T_\text{sort}^b$ is the time spent to sort
a batch of size $b$ and $T_\text{merge}^b$
is the time spent to merge two batches of size $b$.
The second term can be derived because merge is a linear operation, and hence merging two arrays each of size $2^k b$ takes about $2^{k+1}T_\text{merge}$.

Because of this complex behavior and in order to better compare the GPU LSM's performance with GPU SA and hash tables, we conduct the following experiment with a fixed batch size $b$: We randomly generate $n = 2^{27}$ elements and incrementally insert batches of $b$ elements ($\lceil n/b \rceil$ batches) into the GPU LSM\@.
After each insertion, we compute the insertion rate for that batch ($b$ divided by the insertion time).
By doing so, we have considered all possible resident batches of $1 \leq r \leq \lceil n/b \rceil$.
We continue this approach for different batch sizes.

Table~\ref{table:insertion_rate} shows the minimum rate (when all levels are full), the maximum rate (when the first level is empty), and the harmonic mean of all possible other outcomes.
We have also repeated the same experiment for the GPU SA\@.
The minimum rate for GPU LSM is usually worse than the GPU SA's because of iterative merges. However, the mean rate is always better than the GPU SA, because in practice, on average, very few merges are performed before
finding an empty level.
For a fixed $n$, if a smaller $b$ is chosen for the GPU LSM, there will be more full levels and hence more iterative merges and as a result slower performance.
Averaged (harmonic mean) over all batch sizes, the GPU LSM's insertion rate is 225.3~M~elements/s, which is 13.5x faster than SA\@.

\paragraph{Effective insertion rate}
How does the GPU LSM compare with the GPU SA with repeated inserts? The effective insertion rate for the GPU LSM gets increasingly better than the GPU SA as more and more batches are inserted.
The reason is that for the GPU SA to handle insertions, each time the new batch must be sorted and then merged with all other elements ($O(1/n)$ rate).
However, it is possible (by summing and bounding $T_\text{ins}^b(r)$ over a series of batch insertions) to show that for the GPU LSM the effective insertion rate is instead $O(1/\log n)$.
To show this experimentally, we start with
an empty GPU LSM and incrementally insert batches.
After each insertion we compute the effective rate (number of resident elements divided by total time).
Figure~\ref{fig:rate_insertion} shows the results for various batch sizes.
Repeating the same experiment for GPU SA, it is clear that its performance
degrades at a higher rate as the number of elements increases.
In short, while merging into a GPU SA is fast and the GPU SA could potentially be used for dynamic inserts, it is clear both theoretically and in practice that the GPU LSM is the superior data structure for insertions, and, with the tombstoning
scheme, for deletions as well.

\subsection{Queries: Lookup, count and range operations}\label{subsec:perf_queries}
The performance of queries in the GPU LSM depends
on the specific arrangement of full levels in the current
data structure.
Searching more levels requires more work.
Also the larger the levels to be searched,
the more work.
We observe in this section that query performance depends mainly on
the average number of full levels, and hence for a
fixed number of elements,
a GPU LSM with larger batch size usually has superior query performance.

For lookups, we start from the first full level and continue
until we find the query key,
so the worst case happens when the query does not exist in the data structure.
For count and range queries all levels must be searched regardless.
The larger the range in a count or range query (e.g., $k_2-k_1$),
the more elements are expected to fall within it and
hence more work is required for their
final validation, which results in reduced performance.

Because of the complex relation of query performance with level
distributions, we conduct the following experiment: For any fixed batch
size $b$ and total of $n =2^{24}$ elements,
we build every possible GPU LSM with $1\leq r\leq n/b$ resident batches.
In each case, we generate the same number of queries as there are
elements in the data structure.
We report the minimum rate, the maximum rate,
and the harmonic mean of queries.
The same experiment is repeated on a sorted array (GPU SA) of the same size.
\subsubsection{Lookup queries} \label{subsec:perf_lookup}
Table~\ref{table:lookup_rate} shows the result of this experiment
in two scenarios: 1)~all queries exist (right) or 2)~none exist (left).
In this experiment we see that the mean query
rate of the GPU LSM decreases as the batch size decreases,
because on average the total number of occupied levels increases.
Averaged (harmonic mean) over all batch sizes in Table~\ref{table:lookup_rate}, our GPU LSM gets 67.9 (or 75.7) M~queries/s for none (or all) scenarios. Similarly, the GPU SA reaches 119.0 (or 133.3) M~queries/s, which are both almost 1.75x faster than their respective GPU LSM rate. The cuckoo hash table is also 7.34x (or 10.01x) faster than the GPU LSM\@.
Recall, however, that both the GPU SA and hash tables are immutable data structures.
\begin{table}
\scriptsize
    \resizebox{\linewidth}{!}{%
    \begin{tabular}{c cccc|cccc}
    \toprule
    & \multicolumn{4}{c}{none existing} & \multicolumn{4}{c}{all existing} \\
    \cmidrule(lr){2-5} \cmidrule(lr){6-9}
    $b$ & min & max & mean & GPU SA & min & max & mean & GPU SA \\
    \midrule
    $2^{24}$ & 116.8 & 116.8 & 116.8 &   106.8 &  106.8 & 106.8 & 106.8 & 116.7 \\
    $2^{23}$ & 116.8 & 133.1 & 124.4 &  112.4 & 106.8 & 118.6 & 112.4 & 124.3\\
    $2^{22}$ & 61.6 & 183.1 & 105.9 &  113.9& 74.7 & 142.2 & 104.7 & 128.9\\
    $2^{21}$ & 41.4 & 237.4 & 84.0 &  118.4 & 51.7 & 194.7 & 91.1 &  132.9\\
    $2^{20}$ & 30.7 & 291.3 & 68.4 &  120.3 & 39.3 & 251.1 & 77.6 &  135.0\\
    $2^{19}$ & 25.2 & 331.7 & 57.7 &  123.4 & 32.0 & 299.2 & 67.5 &  138.7\\
    $2^{18}$ & 22.0 & 361.5 & 51.1 &  125.4 & 27.2 & 332.0 & 60.7 &  141.2\\
    $2^{17}$ & 19.5 & 376.7 & 47.4 &  126.8 & 24.3 & 347.3 & 56.7 &  143.0\\
    $2^{16}$ & 18.1 & 386.1 & 45.5 &  127.6 & 22.7 & 365.7 & 54.9 &  144.1\\
    \midrule
    \midrule
    \multicolumn{2}{c}{Cuckoo hash} & \multicolumn{3}{c|}{498.9} & \multicolumn{4}{c}{758.0} \\
    \bottomrule
    \end{tabular}
    }
    \caption{\small Lookup rate (M~queries/s): searching for the same number of keys as there are elements in the data structure such that no (left) or all (right) queries exist. For GPU LSM, we have gathered results from all possible number of batches for $1 \leq r \leq 2^{24}/b$. For GPU SA we have only reported the harmonic mean. Cuckoo hashing is with 80\% load factor.}\label{table:lookup_rate}
\end{table}
\subsubsection{Count and Range queries} \label{subsec:perf_count_range}
Unlike lookup queries, count and range queries will always search through all full levels of a GPU LSM\@.
However, there is a new important factor in their performance: the \emph{expected range} ($L$) of each argument.
The larger the $L$, the more keys fall within its range, and hence more potential results must be validated.
Table~\ref{table:range_count_rate} shows the same experiment we describe at the beginning of this section, followed by the same number of queries with two different $L$ values.
Averaged over all batch sizes, count queries for $L = \{8, 1024\}$ reach \{32.7, 2.8\} M~queries/s, which is \{1.84x, 1.45x\} slower than performing the same queries on a SA with the same size.
Similarly for range queries and averaged over all batch sizes, we reach \{23.3, 1.3\} M~queries/s, which is \{1.39x, 1.36x\} slower than the GPU SA\@.
Noted that it is not possible to efficiently use hash tables for any of these queries at all.
Count queries are also always faster than range queries because range queries have to perform a more expensive validation stage (Section~\ref{subsec:range_details}), and in the end they have to return all valid elements as opposed to just sending back a total count.

\begin{table}
\scriptsize
    \resizebox{\linewidth}{!}{%
    \begin{tabular}{lc cccc|cccc}
    \toprule
    & & \multicolumn{4}{c}{$L = 8$} & \multicolumn{4}{c}{$L = 1024$} \\
    \cmidrule(lr){3-6} \cmidrule(lr){7-10}
    & $b$ & min & max & mean & GPU SA & min & max & mean & GPU SA \\
    \midrule
    \multirow{5}{*}{\rotatebox[origin=c]{90}{Count Query}}
    & $2^{20}$ & 27.0 & 103.7 & 48.0 & 73.5 & 2.58 & 4.17 & 3.01 & 4.10 \\
    & $2^{19}$ & 22.4 & 94.1 & 39.8 & 68.6 & 2.51 & 4.13 & 2.86 & 4.09 \\
    & $2^{18}$ & 19.0 & 82.9 & 35.1 & 63.0 & 2.44 & 4.10 & 2.79 & 4.08 \\
    & $2^{17}$ & 16.2 & 67.6 & 27.4 & 56.1 & 2.36 & 4.06 & 2.63 & 4.04 \\
    & $2^{16}$ & 15.1 & 44.7 & 23.9 & 47.3 & 2.32 & 3.99 & 2.57 & 3.99 \\
    \midrule
    \midrule
    \multirow{5}{*}{\rotatebox[origin=c]{90}{Range Query}}
    & $2^{20}$ & 23.2 & 72.1 & 38.4 & 42.2 & 1.27 & 1.80 & 1.43 & 1.81 \\
    & $2^{19}$ & 19.4 & 63.2 & 31.9 & 38.2 & 1.25 & 1.81 & 1.37 & 1.81 \\
    & $2^{18}$ & 14.6 & 50.5 & 24.9 & 33.8 & 1.23 & 1.81 & 1.35 & 1.81 \\
    & $2^{17}$ & 12.9 & 36.3 & 19.8 & 29.1 & 1.20 & 1.81 & 1.28 & 1.81 \\
    & $2^{16}$ & 11.0 & 23.0 & 15.1 & 25.0 & 1.18 & 1.76 & 1.27 & 1.80 \\
    \bottomrule
    \end{tabular}
    }
    \caption{\small Count and Range queries in GPU LSM and GPU SA with expected range $L = 8,1024$. We have gathered results from all possible number of batches for $1 \leq r \leq 2^{24}/b$.}
    \label{table:range_count_rate}
\end{table}

\subsection{Cleanup and its relation with queries}\label{subsec:perf_cleanup}
Our observations show that the cleanup operation---discussed in Section~\ref{subsec:cola_cleanup} and detailed in Section~\ref{subsec:cleanup_details}---is much more efficient than rebuilding the whole data structure from scratch.
Also, it can greatly improve query performance by reducing the number of full levels and hence may be useful to perform regularly.

Our experiments show that the
speed of the
cleanup operation mostly depends on the total number
of resident elements in the data structure, and less significantly
on the percentage of elements that need to be removed
(stale elements).
The more elements to be removed the better.
For example, with $n = (2^6-1)b$ elements where $b = 2^{20}$, cleanup operations when \{10, 50\}\% of elements should be removed runs at \{1870.2, 1828.2\} M~elements/s.
A GPU LSM with roughly the same size ($n = (2^7-1)b$ with $b = 2^{19}$) with \{10, 50\}\% of elements removed results in \{1842.5, 1794.3\} M~elements/s.
Since the GPU LSM's bulk build sustains 728~M~elements/s (Section~\ref{subsec:perf_insertion}), cleanup is up to 2.5x faster than building all elements from scratch.

Depending on the number of resident batches, cleanup can speed up queries
by potentially reducing the total number of full levels.
For example, with 10\% removals, $n = (2^7-1)$, and $b = 2^{18}$, cleanup takes 19.23~ms to finish. After cleanup, we can perform 32 million lookup queries in 132.5~ms, which is almost 4.8x faster than performing the exact same queries before the cleanup (including the cleanup time).
This is an important result, since it motivates the user to regularly perform cleanups if she needs to perform a lot of queries during the lifetime of a GPU LSM\@.

\section{Conclusion}\label{sec:conclusion}
We proposed and implemented the GPU LSM\@,
a dynamic dictionary data structure suitable for GPUs,
with fast batch update operations (insertions and deletions)
as well as a variety of parallel queries (lookup, count, and range).
We find that we can update the GPU LSM much more efficiently than
we can a sorted array, especially for small batch sizes, at the cost of
a small constant factor in query time.
It might be possible to improve the
query time by employing the fractional cascading idea used in
COLA~\cite{Bender:2007:CSB}, at the cost of a more complicated insertion
and more memory; it is not clear whether this would be practical or not.
Similar ideas might be useful in other GPU data structures,
for example BVH trees,
which might be useful for applications such as collision detection and
ray tracing in dynamic scenes.

\section*{Acknowledgments}\label{sec:ack}
Thanks to NVIDIA who provided the GPUs that made this research possible.
We appreciate the funding support from a 2016--17 NVIDIA Graduate Fellowship, from NSF awards CCF-1637442, CCF-1724745, CCF-1715777, CCF-1637458, IIS-1541613, IIS-1117663, and IIS-096435, from an Adobe Data Science Research Award, and gifts from EMC and NetApp.

\let\doi\relax                  
\bibliographystyle{IEEEtran}
\bibliography{all,temp,allpapers} 

\end{document}